\documentclass{jfmnomark}
\usepackage{tabularx}
\usepackage{booktabs}
\usepackage{graphicx}
\usepackage{natbib}
\usepackage{amsmath}
\usepackage{pifont}

\usepackage{mathrsfs}
\usepackage{microtype}

\usepackage{tikz}

\usepackage{tabularx}

\ifCUPmtlplainloaded \else
  \checkfont{eurm10}
  \iffontfound
    \IfFileExists{upmath.sty}
      {\typeout{^^JFound AMS Euler Roman fonts on the system,
                   using the 'upmath' package.^^J}%
       \usepackage{upmath}}
      {\typeout{^^JFound AMS Euler Roman fonts on the system, but you
                   dont seem to have the}%
       \typeout{'upmath' package installed. JFM.cls can take advantage
                 of these fonts,^^Jif you use 'upmath' package.^^J}%
      }
  \else
  \fi
\fi

\ifCUPmtlplainloaded \else
  \checkfont{msam10}
  \iffontfound
    \IfFileExists{amssymb.sty}
      {\typeout{^^JFound AMS Symbol fonts on the system, using the
                'amssymb' package.^^J}%
       \usepackage{amssymb}%
         \let\leq=\leqslant
         \let\geq=\geqslant
      }{}
  \fi
\fi

\ifCUPmtlplainloaded \else
  \IfFileExists{amsbsy.sty}
    {\typeout{^^JFound the 'amsbsy' package on the system, using it.^^J}%
     \usepackage{amsbsy}}
    {}
\fi

\newsavebox{\astrutbox}
\sbox{\astrutbox}{\rule[-5pt]{0pt}{20pt}}

\newcommand\eu{u}
\newcommand{\ol}[1]{\overline{#1}}

\title[Exponential Asymptotics and Ship Waves]{Waveless ships in the low speed limit: \\ Results for multi-cornered hulls}

\author[P.H. Trinh and S.J. Chapman]%
{P\ls H\ls I\ls L\ls I\ls P\ls P\ls E\ns H.\ns T\ls R\ls I\ls N\ls H$^{1, 2}$\ns \and 
S.\ns J\ls O\ls N\ls A\ls T\ls H\ls A\ls N\ns C\ls H\ls A\ls P\ls M\ls A\ls
N$^2$ \break}

\affiliation{ $^1$Program in Applied and Computational Mathematics, Princeton University, \\ Washington Road,
Princeton, NJ, 08544, USA \\[\affilskip]
$^2$Oxford Centre for Industrial and Applied Mathematics, Mathematical Institute, \\ 24-29 St. Giles',
Oxford, Oxfordshire, OX1 3LB, UK}
\pubyear{1996}
\volume{538}
\pagerange{119--126}
\date{--- and in revised form ---}

\def\Xint#1{\mathchoice
   {\XXint\displaystyle\textstyle{#1}}%
   {\XXint\textstyle\scriptstyle{#1}}%
   {\XXint\scriptstyle\scriptscriptstyle{#1}}%
   {\XXint\scriptscriptstyle\scriptscriptstyle{#1}}%
   \!\int}
\def\XXint#1#2#3{{\setbox0=\hbox{$#1{#2#3}{\int}$}
     \vcenter{\hbox{$#2#3$}}\kern-.5\wd0}}

\def\dashint{\Xint-}

\newcounter{subeq}
\renewcommand{\thesubeq}{\theequation\alph{subeq}}
\newcommand{\newsubeqblock}{\setcounter{subeq}{0}\refstepcounter{equation}}
\newcommand{\mysubeq}{\refstepcounter{subeq}\tag{\thesubeq}}

\newcommand{\minus}{-}

\begin{document}

\maketitle

\begin{abstract}
In the low-speed limit, a blunt ship modeled as two-dimensional semi-infinite body with a single corner \emph{can never be made waveless}. This was the conclusion of the previous part of our work in \cite{trinh_1hull}, which focused on the \cite{dagan_1972} model of ship waves in the low speed limit. In this accompanying paper, we continue our investigations with the study of more general, piecewise-linear, or multi-cornered ships. The low-speed or low-Froude limit, coupled with techniques in exponential asymptotics allows us to derive explicit formulae relating the geometry of the hull to the form of the waves. Configurations with closely spaced corners present a non-trivial extension of the theory, and we present the general methodology for their study. Lastly, numerical computations of the nonlinear ship-wave problem are presented in order to confirm the analytical predictions. 
\end{abstract}


\section{Introduction}

%Is it demonstrable\ldots that continuous solutions will not exist in the limit of vanishing speed? Does this have anything to do with the inability of Tuck and his colleagues in Adelaide to find a continuous solution in the two-dimensional bow wave case?

\noindent The investigations in this paper are focused on the analysis of the low-speed, or low-Froude\footnote{The Froude (draft) number represents the ratio between inertial and gravitational forces.}, wave models proposed by \cite{dagan_1969, dagan_1972}, in which blunt-bodied ships are studied in the context of potential flow and asymptotic expansions in powers of the Froude number. As a particularly interesting case that draws our attention, we recall the work of \cite{farrow_1995}, who showed that by attaching a bulbous-like obstruction to an otherwise rectangular ship's stern, one could produce a dramatic effect on the production of transverse waves. As they reported in their paper:
\begin{quotation}
\noindent \emph{At this [Froude number], a rectangular stern generates waves with steepness $0.0855$, whereas the stern with the downward-pointing bulb [\ldots] yields waves with steepness $0.0119$. It is clear that the addition of the downward-pointing bulb has had a dramatic effect on the downstream wave steepness, reducing it by a factor of 7.2, although it has still not eliminated the waves entirely.}
\end{quotation}

\noindent Our goal is to give an analytical criterion that explains why this phenomenon occurs; that is to say, what distinguishes the two ships, one with a bulb and one without a bulb, in the context of the `slow-ship' approximation? The advantage of the slow-ship potential-flow approximation, is that allows us to directly relate the generation of waves to the shape of the ship's hull without the need for numerical simulations. 

In addition to addressing the \cite{farrow_1995} issue, we are also interested in a more general question: in the low-speed limit, when a blunt ship is modeled as a two-dimensional semi-infinite body, can it ever be made waveless? These waveless or wave-minimisation questions in the context of the \cite{dagan_1972} approximation were studied by \cite{vb_1977}, \cite{vb_1978}, \cite{madurasinghe_1986}, and \cite{tuck_1991, tuck_1991b} for ship hulls of varying geometries, and we are interested in continuing their line of inquiry.

\begin{figure}
\includegraphics[width=1.0\linewidth]{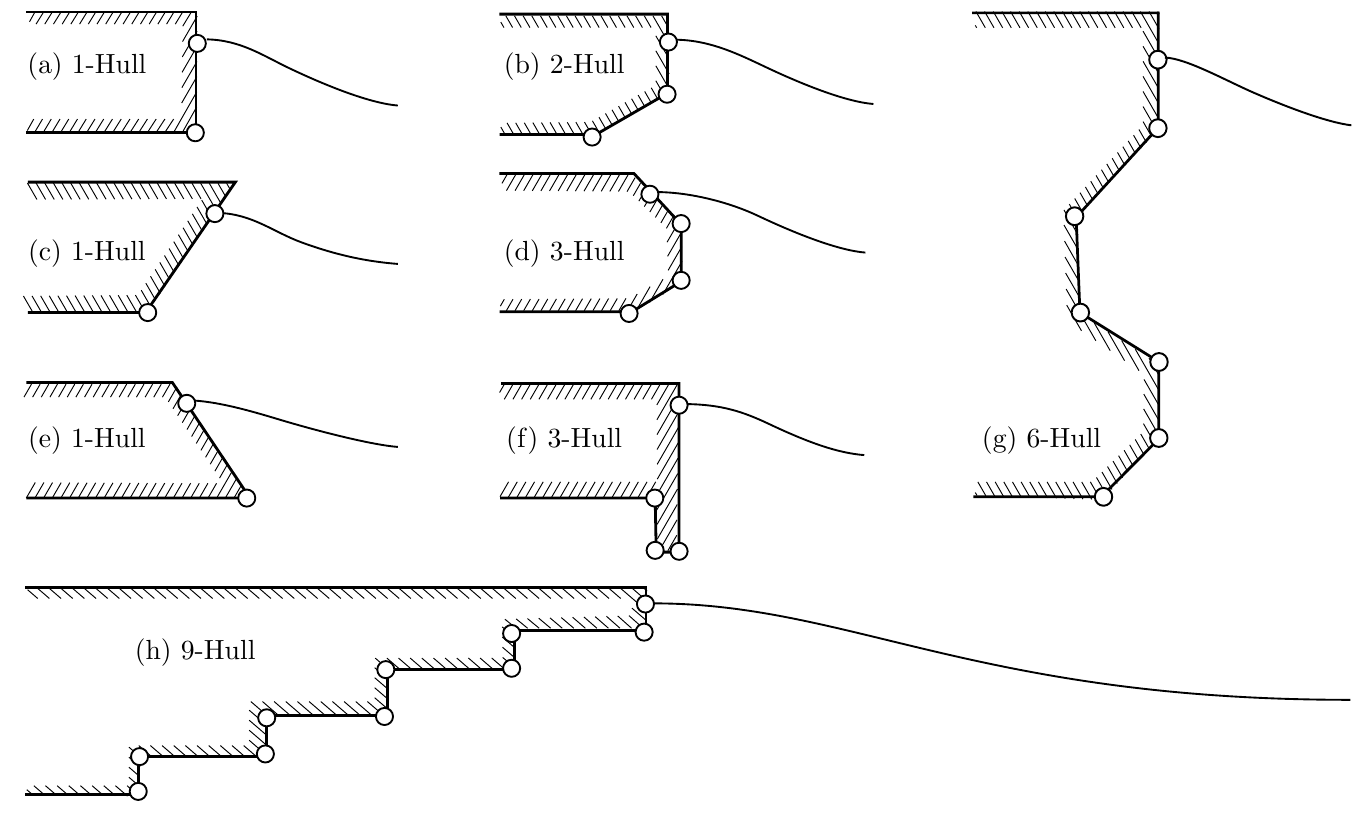}
% \beginpgfgraphicnamed{nhull_guess}
% \begin{tikzpicture}
% \node at (0,0){\includegraphics[width=1.0\linewidth]{figpdf/guess.pdf}};
% \node[scale=0.8] at (-6.1,3.45) {(a) 1-hull};
% \node[scale=0.8] at ( -1.25,3.45) {(b) 2-hull};
% 
% \node[scale=0.8] at (-6.1,1.55) {(c) 1-hull};
% \node[scale=0.8] at ( -1.25,1.55) {(d) 3-hull};
% 
% \node[scale=0.8] at (-6.1,-0.35) {(e) 1-hull};
% \node[scale=0.8] at ( -1.25,-0.35) {(f) 3-hull};
% \node[scale=0.8] at ( 3.6,-0.35) {(g) 6-hull};
% 
% \node[scale=0.8] at (-5.0,-2.5) {(h) 9-hull};
% \end{tikzpicture}
% \endpgfgraphicnamed
\caption{Are any of these ships waveless? In all cases, the flow is from left
to right and nodes indicate singularities in the analytic continuation. \label{fig:nhull_ships}}
\end{figure}

In the previous part of our work (\citealt{trinh_1hull}), we demonstrated that piecewise-linear hulls with a single, submerged corner can never be made waveless\footnote{Consequently, a free surface that attaches to a single-cornered bow at a stagnation point is not possible within the \cite{dagan_1972} model.}. For the case of piecewise-linear hulls with multiple corners, the answer to this question is not as clear. For example, are any of the eight hulls presented in Figure \ref{fig:nhull_ships} waveless? If not, then which ones produce the smallest waves? For the case of potential flow over a submerged obstruction, waveless configurations are certainly possible, as was demonstrated by \cite{lustri_2012} and \cite{hockingGC_2012}, but the same question for surface-piercing ships of general form remains open. Certainly, there are notable difficulties in studying this problem. For example, waveless ships were proposed by \cite{tuck_1984} and \cite{madurasinghe_1986}, but these were later refuted in the more comprehensive numerical study by \cite{farrow_1995}, in which they showed that
\begin{quotation}
\noindent \emph{The free surface would at first sight appear to be waveless, but on closer examination of the numerical data, there are very small waves present and they have a steepness of $1.5 \times 10^{-3}$}
\end{quotation}

\noindent in reference to the bulbous hull in \cite{tuck_1984}. Our desire, then, is to study these issues in terms of the low-Froude asymptotic expansions. 

Having progressed through the theory of \cite{trinh_1hull}, we know that at low Froude numbers, the waves generated by a ship become exponentially small and are thus invisible to a regular asymptotic expansion. The ineffectiveness of traditional asymptotics in capturing the low-speed limit was first remarked by \cite{ogilvie_1968, ogilvie_1970} and later termed the \emph{Low-Speed Paradox}. Techniques in exponential asymptotics \citep{boyd_wnlsw} allow us to demonstrate the fact that these hidden waves are switched-on when the regular expansion is continued across critical curves (\emph{Stokes lines}) in the complex plane; this process is known as the \emph{Stokes Phenomenon}. Most important, however, is the valuable insight that these approximations give: an explicit formula that relates the shape of an arbitrary hull with its resultant waves. The question of wavelessness in the low-Froude limit is then simplified to examining whether the sum of the Stokes line contributions can ever be zero in regions far from the ship. 

The requisite background in exponential asymptotics can be found in \cite{trinh_1hull}. The techniques we apply are based on the use of a factorial over power ansatz to capture the divergence of the asymptotic expansions, then optimal truncation and Stokes line smoothing to relate the late-order terms to the exponentially small waves [see for example, papers by \cite{olde_1995}, \cite{chapman_1998}, and \cite{trinh_2010}]. Our paper also parallels the works of \cite{chapman_2002, chapman_2006} and \cite{ trinh_gclinear, trinh_gcnonlinear} on the application of exponential asymptotics to the study of gravity or capillary waves produced by flow over a submerged object.

\subsection{The role of low-Froude approximations and an outline of the paper}

It is important for us to mention that the low-Froude model of \cite{dagan_1972} is indeed a very \emph{idealised} approximation for understanding the production of ship waves. Real stern and bow flows are very complex, and viscosity, turbulence, and necklace vorticies can all play an important role in the production of waves. We refer the reader to, for example, some of the numerical simulations of \cite{grosenbaugh_1989} and \cite{yeung_1997} that demonstrate some of the complex dynamics that arise in ship flows once, for example, vorticity and viscosity are included. In \S\ref{sec:discuss} of this paper, we shall return to discuss the caveats of the low-speed approximation.

%Our work shows how exponential asymptotics can be used for such flows. So far we have only considered inviscid models but we are now considering extensions of exponential asymptotics techniques to more complex situations (gravity capillary flows, three-dimensional effects, etc).

Ultimately, we are interested in obtaining analytical intuition about the connection between the ship's hull and the waves produced. The more usual routes towards analytical solutions assumes an asymptotically small geometry, which leads to the `thin-ship', `flat-ship' or `streamline-ship' approximations; in such regimes, a waveless ship is impossible (see for example, \citealt{kotik_1964} and Krein in \citealt{kostyukov_1968}), but these theories say very little about the case of non-thin ships. Other examples of ship wave models can be studied, including the Kelvin-Neumann formulation in which the free surface condition is linearised about a steady uniform stream annd the boundary condition on the ship's hull is satisfied exactly, but generally these problems do require a degree of numerical computation. A discussion of such problems can be found in the book by \cite{kuznetsov_book} and the review and discussion by \cite{newman_1991} (see \emph{e.g.} \citealt{pagani_2004} for more recent work on rigorous results applicable to non-slender geometries). We finally refer readers to the reviews by \cite{tuck_1991} and \cite{tulin_2005} for a summary of the role played by low-Froude approximations, particularly in connection with problems in which we require asymptotic solutions that preserve the nonlinearity of the geometry.

% In this paper, we study the more open-ended problem of waves produced by a two-dimensional semi-infinite ship with a piecewise-linear hull. Little, if any, literature exists on the topic, perhaps for the simple reason that without analytical guidelines, one has great difficulty distinguishing between, for example, the wake of a three-cornered ship and that of a five-cornered ship. Numerical results can always be proffered, but the infinity of possible hull configurations makes it rather difficult to infer from quantitative data. The only mention of such piecewise-linear hulls can be found in \cite{farrow_1995}, for which a certain three-cornered ship is briefly studied. 

% The goal of this work is not to provide an all-inclusive theory which encompasses the most general of multi-cornered ships, but rather, to provide an analytical roadmap for understanding how the waves produced by a ship can be directly related to its chosen geometry. In the end, we would like the ability to scan each of the eight hulls presented in Figure \ref{fig:nhull_ships} and answer the question: \emph{are any of these ships waveless?}  

The paper will proceed as follows. First, the mathematical formulation of the ship-wave problem is briefly recapitulated in \S\ref{sec:nhull_form}. This is followed by the asymptotic analysis of the low-Froude problem in \S\ref{sec:nhull_asymptotic}, which culminates with the derivation of explicit expressions for the wake of an arbitrary multi-cornered ship. From these analytical results, we explain why certain classes of multi-cornered ships can never be made waveless in \S\ref{sec:nhull_nonexistence}, then 
in \S\ref{sec:nhull_numerical}, these theoretical results are vindicated by comparisons with numerical computations. 

\section{Mathematical formulation} \label{sec:nhull_form}

\begin{figure} \centering
\includegraphics[width=1.0\linewidth]{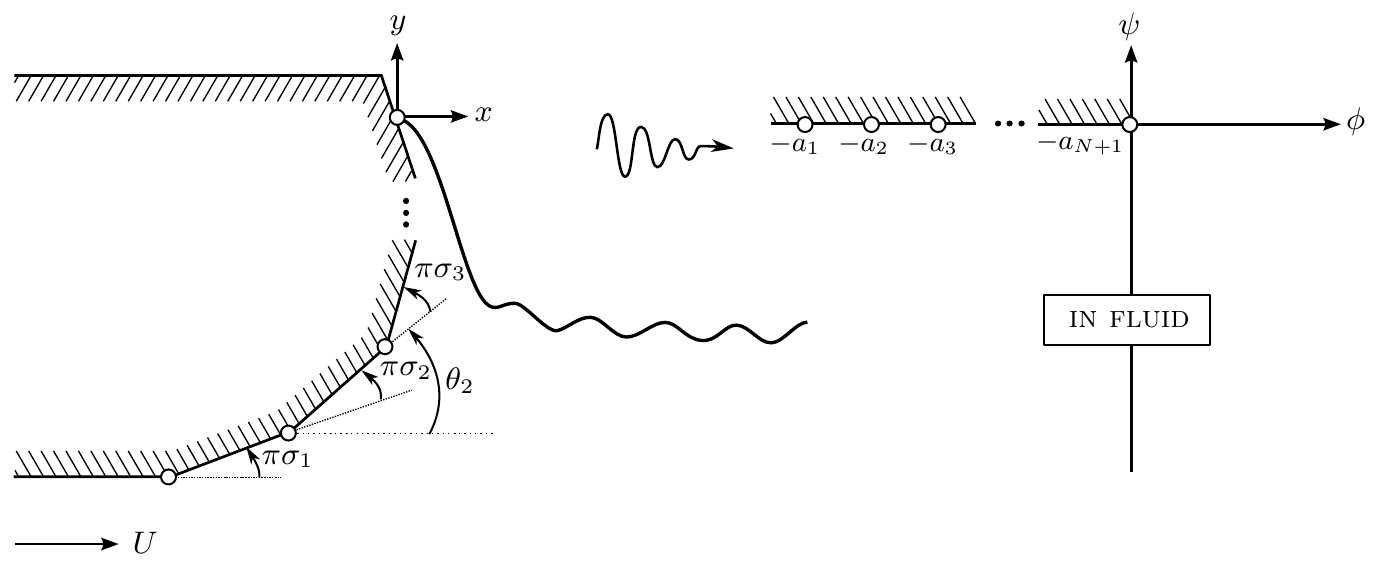}
%\begin{preview}
% \begin{tikzpicture}
% \node at (0,0){\includegraphics[width=1.0\linewidth]{figpdf/form.pdf}};
% \node at (-2.83, 2.75){$y$};
% \node at (-1.96, 1.85){$x$};
% 
% \node[scale=1.0] at (-3.95,-1.65){$\pi \sigma_1$};
% \node[scale=1.0] at (-2.75,-0.75){$\pi \sigma_2$};
% \node[scale=1.0] at (-2.20,-0.85){$\theta_2$};
% \node[scale=1.0] at (-2.4, 0.25){$\pi \sigma_3$};
% 
% \node at (4.6,2.75){$\psi$};
% \node at (6.9,1.78){$\phi$};
% 
% \node[scale=0.9] at (1.2,1.55){$-a_1$};
% \node[scale=0.9] at (1.9,1.55){$-a_2$};
% \node[scale=0.9] at (2.6,1.55){$-a_3$};
% \node[scale=0.9] at (4.10,1.55){$-a_{N+1}$};
% 
% \node at (-5.4,-2.5){$U$};
% \node at (4.6,-0.23){\footnotesize \scshape in fluid};
% \end{tikzpicture}
%\end{preview}
\caption{Flow past a piecewise-linear $N$-hull. The $N$ corners of exterior
angles $\pi\sigma_1, \pi\sigma_2, \ldots, \pi\sigma_N$ in the $(x,y)$
plane (left) are mapped to $w = -a_1, -a_2, \ldots, -a_N$ in the complex
potential plane (right). The stagnation point is $w = -a_{N+1} = 0$. \label{fig:nhull_Nhull}}
\end{figure}

\noindent Consider steady, incompressible, irrotational, inviscid flow in the presence of gravity, past the semi-infinite body shown in Figure \ref{fig:nhull_Nhull}, which consists of a flat bottom and a piecewise linear front face. There is a uniform stream of speed $U$ as $x \to -\infty$, and we assume that the flow attaches to stern\footnote{The reversible nature of potential flow implies that any stern flow can be reversed to bow flow, with the additional condition that there are no waves far upstream from the ship (the radiation condition).} at a stagnation point, $x = 0$ and $y = 0$. 

The dimensional problem can be reposed in terms of a non-dimensional boundary-integral formulation in the potential $(\phi, \psi)$-plane. The unknowns are the fluid speed $q = q(\phi, \psi)$, and streamline angles, $\theta = \theta(\phi, \psi)$, measured from the positive $x$-axis. The body and free-surface are given by the streamline $\psi = 0$, and we assume the free-surface ($\phi > 0$) attaches to the hull ($\phi < 0$) at $\phi = 0$. The free-surface, with $\psi = 0$, is then obtained by solving a boundary-integral equation, coupled with Bernoulli's condition:
\begin{subequations} \label{eq:nhulleqtogether}
\begin{alignat}{2}
 \log{q} &= \frac{1}{\pi} \dashint_{-\infty}^{\infty}
\frac{\theta(\varphi)}{\varphi - \phi} \ d\varphi \label{eq:nhull_bdint} \\
 \epsilon q^2 \frac{dq}{d\phi} &= -\sin{\theta},
\label{eq:nhull_bern}
\end{alignat}
\end{subequations}

\noindent where $\epsilon = U^3/gK$ is related to the square of the Froude draft number with upstream flow $U$, and $K$ is defined by $K = \sum_{i=1}^N K_i$, where $N$ is the number of corners and $\phi_i^* = -K_i$ is the dimensional value of the potential at each of the corners. In this way, if $\phi = a_i$ for $1 \leq i \leq N$ denotes the value of the potential at the corners in the non-dimensionalised 
problem, we have the property that $\sum_{i=1}^N a_i = 1$. The derivation of (\ref{eq:nhulleqtogether}) can be found in \cite{trinh_1hull}, and the only difference is the choice of scaling for the Froude number.

We shall refer to the $N$-cornered piecewise-linear hull as an $N$-hull. For $\phi<0$ the geometry of the hull can be described by 
\begin{equation} \label{thetasig}
\theta(\phi) = \theta_k = \pi \sum_{j=1}^k \sigma_j, 
\end{equation}

\noindent for $a_k < \phi < a_{k+1}$ where $k = 1, \ldots, N$,
$a_{N+1}=0$ is the stagnation point, and $\pi \sigma_k$ is the
exterior angle at the $k$th corner (see Figure \ref{fig:nhull_Nhull}).
When $N = 2$, we will sometimes also refer to the ship as a $[\sigma_1,
\sigma_2]$-hull.  

In (\ref{eq:nhull_bdint}), we write the portion of the boundary integral over the negative
real axis as 
\begin{equation} \label{eq:nhull_rigid}
 \frac{1}{\pi} \dashint_{-\infty}^{0}
\frac{\theta(\varphi)}{\varphi - \phi} \ d\varphi = \log\left[\prod_{k=1}^{N+1}
(\phi+a_k)^{-\sigma_k}\right] \equiv \log q_s(\phi),
\end{equation}

\noindent where $\sigma_{N+1}$ is defined according to the requirement that the free-surface approaches the uniform stream, with $\theta \to 0$ and $q \to 1$ as $\phi \to \infty$; this gives
\begin{equation} \label{eq:nhull_sigtot}
\sigma_{N+1} = -\sum_{j=1}^{N} \sigma_j.
\end{equation}

\noindent The function $q_s$ in (\ref{eq:nhull_rigid}) serves to distinguish the different sorts of piecewise-linear ships. Note also that the product representation of the complex quantity $q_s$ can be alternately derived by using a Schwartz-Christoffel mapping applied to the polygonal hull shape and a rigid, flat free surface. 

As explained in \cite{trinh_1hull}, in order to study the Stokes Phenomenon, we must `complexify' the free-surface, and thus allow $q(\phi, 0) \mapsto q(w)$ and $\theta(\phi, 0) \mapsto \theta(w)$ to be complex functions of the complex potential, $\phi + i0 \mapsto w$. Analytically continuing (\ref{eq:nhull_bdint}) and (\ref{eq:nhull_bern}) gives
\begin{subequations}
\begin{eqnarray}
 \log{q} \pm i\theta &=& \log q_s(w) + \mathscr{H}\theta(w)
\label{eq:nhull_bdint2}
\\
 \epsilon q^2 \frac{dq}{dw} &=& -\sin{\theta} \label{eq:nhull_bern2},
\end{eqnarray}
\end{subequations}

\noindent where the $\pm$ signs correspond to analytic continuation in the
upper and lower-half $\phi$ planes, respectively, and $\mathscr{H}$
denotes the Hilbert Transform operator on the free-surface,
\[
\mathscr{H}\theta(w) = \frac{1}{\pi} \int_0^\infty \frac{\theta(\varphi)}{\varphi-w} \
d\varphi.
\]

%\noindent In the following asymptotic analysis, we will analytically continue
%into the upper-half plane and find exponentially small terms switched on across
%Stokes lines. However, once the analysis is complete, a similar procedure must
%be performed in the lower-half plane as well.

%Because the
%\emph{zero}-Froude solution requires a horizontal attachment angle between the
%free surface and the hull, this also implies that
%\begin{equation} \label{eq:nhull_sigtot}
%\theta_{N+1} = 0 \quad \text{or} \quad \sigma_{N+1} = -\sum_{j=1}^{N+1}
%\sigma_j.
%\end{equation}

\section{Exponential asymptotics} \label{sec:nhull_asymptotic}

\noindent A single-cornered ship will always produce exponentially small waves in the low Froude limit, $\epsilon \to 0$; these waves are explained by the presence of a Stokes line which emerges from the singularity at the corner. For a multi-cornered ship, the analysis proceeds almost identically to \cite{trinh_1hull}, except now, each corner of the hull has the potential to produce Stokes lines and its own separate wave contribution. In this section, we shall briefly re-apply the methodology of the previous work, and provide the corresponding formulae for the case of an $N$-hull.

\subsection{Late-order terms} \label{sec:nhull_late}

\noindent Here, we perform the asymptotic analysis which corresponds to analytically continuing the free-surface into the upper-half $\phi$-plane; continuation into the lower-half plane produces a complex conjugate contribution, which we add to our results, later. 

We begin by substituting the usual asymptotic expansions
\begin{equation} \label{eq:nhull_asympt}
 \theta = \sum_{n=0}^\infty \epsilon^n \theta_n 
 \text{\ \quad and \ \quad}
 q = \sum_{n=0}^\infty \epsilon^n q_n,
\end{equation}

\noindent into (\ref{eq:nhull_bdint2}) and (\ref{eq:nhull_bern2}) (with the $+$ sign). In the limit $\epsilon \to 0$, the leading-order solution is the rigid-wall flow of (\ref{eq:nhull_rigid}),
\begin{align}
\newsubeqblock
\mysubeq \theta_0 &= 0, \\
\mysubeq q_0 &= q_s = \prod_{k=1}^{N+1} (w+a_k)^{-\sigma_k}, \label{eq:nhull_rigid2} \\
\intertext{while the $\mathcal{O}(\epsilon)$ terms are} 
\newsubeqblock
\mysubeq \theta_1 &= - q_0^2 \frac{dq_0}{dw}, \\
\mysubeq q_1 &= iq_0^3 \frac{dq_0}{dw} + q_0 \mathscr{H}\theta_1(w). \label{eq:nhull_q1}
\end{align}

\noindent Notice that the leading-order solution, $q_0$ in
(\ref{eq:nhull_rigid2}), possesses a singularity at each of the
corners, $w = -a_k$. However, the solution at each subsequent order
involves a derivative of the previous order, so we would thus expect
that as $n \to \infty$, the power of the singularity grows, and the
asymptotic expansions (\ref{eq:nhull_asympt}) exhibit factorial over
power divergence: 
\begin{equation} \label{eq:nhull_ansatz}
 \theta_n \sim \sum_{k=1}^{N+1} \frac{\Theta_k
\Gamma(n+\gamma_k)}{\chi_k^{n+\gamma_k}} 
\text{\ \quad and \ \quad}
 q_n \sim \sum_{k=1}^{N+1} \frac{Q_k
\Gamma(n+\gamma_k)}{\chi_k^{n+\gamma_k}}, 
\end{equation}

\noindent where $\gamma_k$ is complex constant, $Q_k$ and $\chi_k$ are functions
of the complex potential $w$, and $\chi_k(-a_k) = 0$.  For most of the
analysis, however, we can simply choose one of the corners of interest
and add the individual contributions at the end. 

The singularities, $w = -a_k$, are located off the free surface, where the Hilbert Transform
in (\ref{eq:nhull_bdint2}) is evaluated and so, as justified in \cite{trinh_1hull}, $\mathscr{H}\theta_n(w)$ is exponentially subdominant to the terms on the left-hand side for large $n$. At $\mathcal{O}(\epsilon^n)$, (\ref{eq:nhull_bdint2}) gives
\begin{equation} \label{eq:nhull_oen_bdint2}
 \theta_n \sim i\frac{q_n}{q_0} - \frac{iq_1 q_{n-1}}{q_0^2} + \ldots
\end{equation}

\noindent as $n\to \infty$, and substitution into (\ref{eq:nhull_bern2}) gives the relevant terms at $\mathcal{O}(\epsilon^n)$:
\begin{equation}
\biggl[q_0^3 q'_{n-1} + iq_n\biggr] +
\biggl[ 2q_0^2q_0'q_{n-1} + 2q_0^2q_1q_{n-2}' - i\frac{q_{n-1} q_1}{q_0}\biggr] + \ldots = 0.
\label{eq:nhull_oen}
\end{equation}

\noindent We substitute the ansatzes of (\ref{eq:nhull_ansatz}) into (\ref{eq:nhull_oen}), and this yields, at leading order as $n\to\infty$,
\begin{equation} \label{eq:nhull_chiw}
\frac{d\chi}{dw} = \frac{i}{q_0^3}.
\end{equation}

\noindent Using $\chi_k(-a_k) = 0$, we integrate this result, to give 
\begin{equation} \label{eq:nhull_chi}
 \chi_k(w) = \int_{-a_k}^w \frac{i}{q_0^3(\varphi)} \ d\varphi.
\end{equation}

\noindent At the next order as $n \to \infty$, we find that
\begin{equation} \label{eq:nhull_Q}
 Q_k(w) = \frac{\Lambda_k}{q_0^2} \exp\left[ 3i \int_{w^\bigstar}^w
\frac{q_1(\varphi)}{q_0^4(\varphi)} \ d\varphi \right],
\end{equation}

\noindent where $\Lambda_k$ is constant, and $w^\bigstar$ is any point for which the integral is defined. Finally, (\ref{eq:nhull_oen_bdint2}) allows us to relate $Q_k$ with $\Theta_k$, using $\Theta_k \sim iQ_k/q_0$, so that
\[
 \Theta_k(w) = \frac{\Lambda_k i}{q_0^3} \exp\left[ 3i \int_{w^\bigstar}^w
\frac{q_1(\varphi)}{q_0^4(\varphi)} \ d\varphi \right].
\]

\noindent With $\chi_k$, $Q_k$, and $\Theta_k$ now determined, we have thus derived the late-orders behaviour in (\ref{eq:nhull_ansatz}), subject to the values of $\gamma_k$ and $\Lambda_k$; these must be determined by applying the method of matched asymptotics near the singularity, $w = -a_k$. 

\subsection{Stokes lines and the Stokes Phenomenon} \label{sec:nhull_stokes}

\noindent From \cite{trinh_1hull}, we know that the late-order terms (\ref{eq:nhull_ansatz}) play a crucial role in determining the free-surface waves. Using the expression of $\chi_k$ in (\ref{eq:nhull_chi}), Stokes lines can be traced from each of the ship's corners, across which the Stokes Phenomenon necessitates the switching-on of waves.  From \cite{dingle_book}, these special lines are given by the points $w\in\mathbb{C}$ where
\[
 \Im[\chi_k(w)] = 0 \text{\quad and \quad} \Re[\chi_k(w)] \geq 0.
\]

\noindent If we write $q_0 \sim c_k (w+a_k)^{-\sigma_k}$ near $w = -a_k$, then from (\ref{eq:nhull_rigid2}), we have
\begin{equation} \label{eq:nhull_ck}
 c_k = \prod_{\substack{{j=1} \\ j \neq k}}^{N+1} 
(a_j-a_k)^{-\sigma_j},
\end{equation}

\noindent and thus from (\ref{eq:nhull_chi}), in the limit that $w \to -a_k$, we have
\[
 \chi_k \sim \left[\frac{i}{c_k^3(1+3\sigma_k)}\right] (w+a_k)^{1+3\sigma_k}.
\]

\noindent The condition that $\chi_k(-a_k) = 0$ thus requires that $\sigma_k > -1/3$. In other words, for there to be a singularity, the local deviation
of the corner must be greater than $-\pi/3$. This is a \emph{necessary} (but not
\emph{sufficient}) condition for there to be a free-surface wave produced by
the corner. In fact, a stronger condition for the existence of a Stokes line emerging on the
relevant Riemann sheet can be derived. Since $qe^{-i\theta} = u - iv$, we can
write
\begin{equation}
 \arg(c_k) = \theta_k \label{eq:argck}
\end{equation}

\noindent for analytic continuation into the upper-half plane, where
$\theta_k\in (-\pi, \pi)$ is the angle of the hull as $w \to
a_k^{+}$, measured from the positive $x$-axis (shown in Figure
\ref{fig:nhull_Nhull}).
If we write $\arg(w+a_k) = \nu_k$, then Stokes lines must leave at angles of
\begin{equation} \label{eq:nhull_stokescond}
 \nu_k = \left(\frac{3\theta_k + 2m\pi - \pi/2}{1+3\sigma_k}\right),
\end{equation}

\noindent for $m\in \mathbb{Z}$ and we thus need $\nu \in (0, \pi)$ in order for
the line to emerge in the upper-half plane. The general requirements for
a Stokes line to intersect the free-surface is a global function of the
leading-order flow, but for most hulls, the requirement that $\nu \in (0, \pi)$ with (\ref{eq:nhull_stokescond})
is adequate. In any case, we will let $\mathcal{J} \subseteq  \{1,2,\ldots,N+1\}$ denote those corners which have Stokes lines crossing the free-surface. 

\begin{figure}
\includegraphics[width=1.0\linewidth]{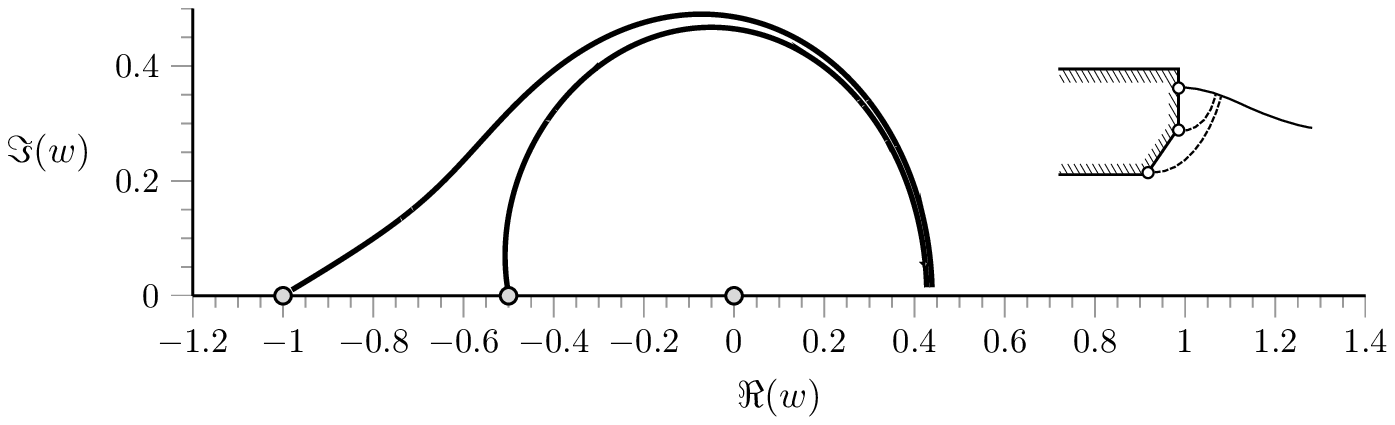}
% \beginpgfgraphicnamed{nhull_stokes1}
% \begin{tikzpicture}  
%   \begin{axis}[ xlabel=$\Re(w)$, 
% 		ylabel=$\Im(w)$, 
% 		ylabel style={font=\large},
% 		xlabel style={font=\large},
% 		ymin=0, ymax=0.5,
% 		xmin=-1.2, xmax=1.4,
% 		%xtick={0.0}, ytick=\empty,
% 		%major tick length=0.22cm,
% 		%minor tick length=0.0cm,
% 		width=\textwidth, height=4.5cm]
% 		
% 	\addplot[smooth, line width=1.5pt] file {plotdata/stokes_2hulla.dat};
% 	\addplot[smooth, line width=1.5pt] file {plotdata/stokes_2hullb.dat};
% 	\addplot[mark=*, only marks, %
% 		mark options={%
% 		scale=1.2, fill=gray!30, draw=black, line width=0.8pt}] 
% 		coordinates {(-1,0) (-0.5,0) (0,0)};
% 	\node at (axis cs:1,0.3)
% 	{\includegraphics[width=2.6cm]{figpdf/mini_2hull.pdf}};	
%   \end{axis}
% \end{tikzpicture}
% \endpgfgraphicnamed

\includegraphics[width=1.0\linewidth]{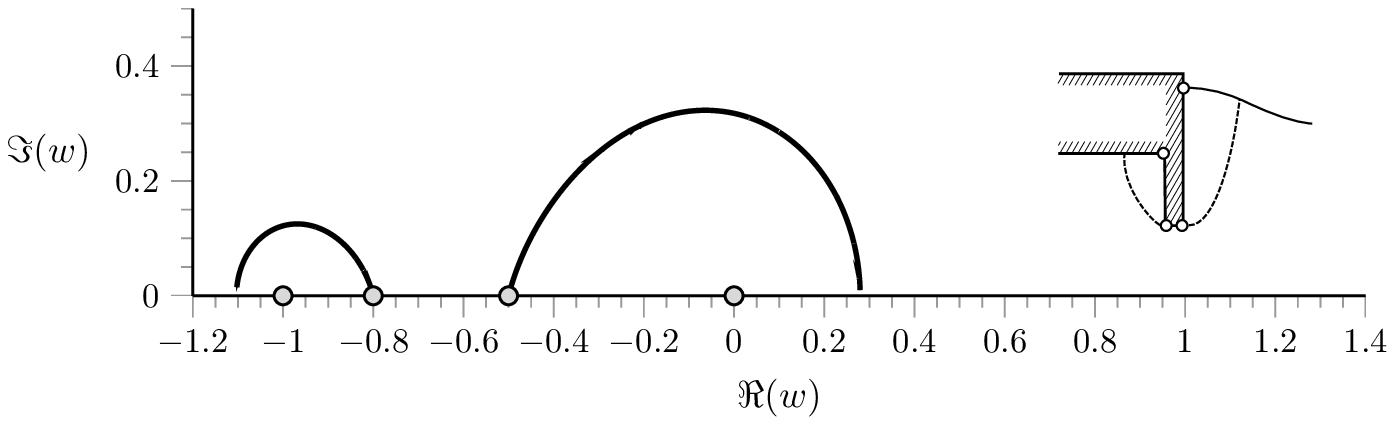}
% \beginpgfgraphicnamed{nhull_stokes2}
% \begin{tikzpicture}  
%   \begin{axis}[ xlabel=$\Re(w)$, 
% 		ylabel=$\Im(w)$, 
% 		ylabel style={font=\large},
% 		xlabel style={font=\large},
% 		ymin=0, ymax=0.5,
% 		xmin=-1.2, xmax=1.4,
% 		%xtick={0.0}, ytick=\empty,
% 		%major tick length=0.22cm,
% 		%minor tick length=0.0cm,
% 		width=\textwidth, height=4.5cm]
% 		
% 	\addplot[smooth, line width=1.5pt] file {plotdata/stokes_farrowa.dat};
% 	\addplot[smooth, line width=1.5pt] file {plotdata/stokes_farrowb.dat};
% 	\addplot[mark=*, only marks, %
% 		mark options={%
% 		scale=1.3, fill=gray!30, draw=black, line width=0.8pt}] 
% 		coordinates {(-1.0,0) (-0.8,0) (-0.5,0) (0,0)};
% 	\node at (axis cs:1,0.25)
% 	{\includegraphics[width=2.6cm]{figpdf/mini_farrowhull.pdf}};	
%   \end{axis}
% \end{tikzpicture}
% \endpgfgraphicnamed

\includegraphics[width=1.0\linewidth]{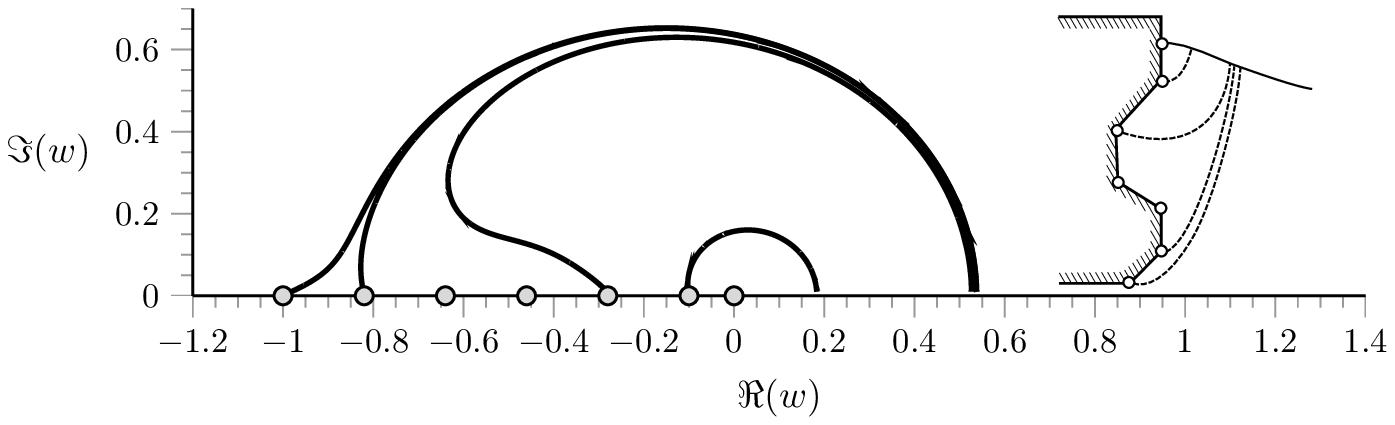}
% \beginpgfgraphicnamed{nhull_stokes3}
% \begin{tikzpicture}  
%   \begin{axis}[ xlabel=$\Re(w)$, 
% 		ylabel=$\Im(w)$, 
% 		ylabel style={font=\large},
% 		xlabel style={font=\large},
% 		ymin=0, ymax=0.7,
% 		xmin=-1.2, xmax=1.4,
% 		%xtick={0.0}, ytick=\empty,
% 		%major tick length=0.22cm,
% 		%minor tick length=0.0cm,
% 		width=\textwidth, height=4.5cm]
% 		
% 	\addplot[smooth, line width=1.5pt] file {plotdata/stokes_6hulla.dat};
% 	\addplot[smooth, line width=1.5pt] file {plotdata/stokes_6hullb.dat};
% 	\addplot[smooth, line width=1.5pt] file {plotdata/stokes_6hulle.dat};
% 	\addplot[smooth, line width=1.5pt] file {plotdata/stokes_6hullf.dat};
% 	\addplot[mark=*, only marks, %
% 		mark options={%
% 		scale=1.3, fill=gray!30, draw=black, line width=0.8pt}] 
% 		coordinates {%
% 		(-1,0) (-0.82,0)
% 		(-0.64,0) (-0.46,0)
% 		(-0.28,0) (-0.1,0) (0,0)};
% 	\node at (axis cs:1,0.35)
% 	{\includegraphics[width=2.6cm]{figpdf/mini_6hull.pdf}};	
%   \end{axis}
% \end{tikzpicture}
% \endpgfgraphicnamed

\includegraphics[width=1.0\linewidth]{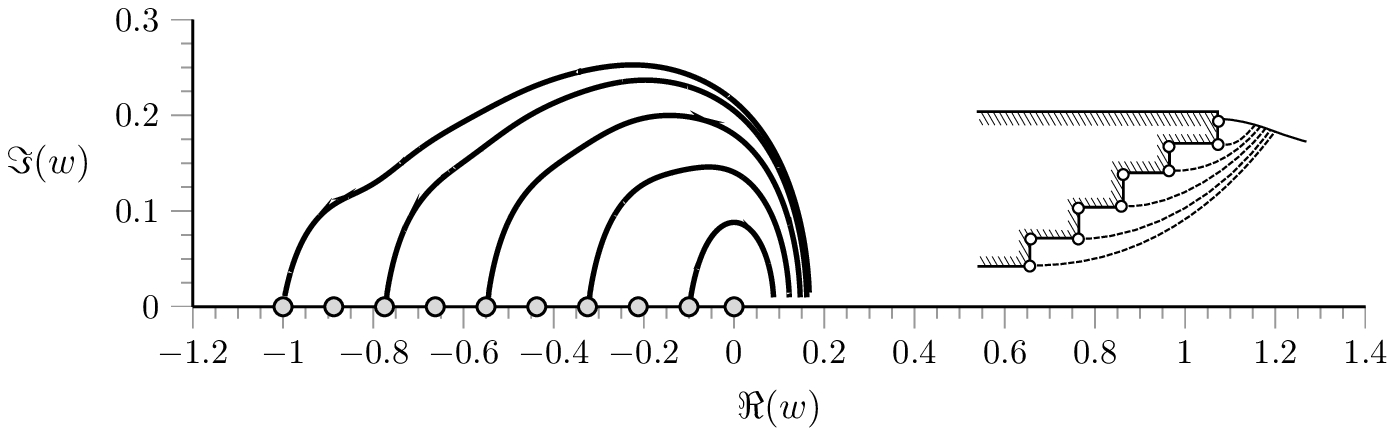}
% \beginpgfgraphicnamed{nhull_stokes4}
% \begin{tikzpicture}  
%   \begin{axis}[ xlabel=$\Re(w)$, 
% 		ylabel=$\Im(w)$, 
% 		%ylabel style={font=\large},
% 		%xlabel style={font=\large},
% 		ymin=0, ymax=0.3,
% 		xmin=-1.2, xmax=1.4,
% 		%xtick={0.0}, ytick=\empty,
% 		%major tick length=0.22cm,
% 		%minor tick length=0.0cm,
% 		width=\textwidth, height=4.5cm]
% 		
% 	\addplot[smooth, line width=1.5pt] file {plotdata/stokes_9hulla.dat};
% 	\addplot[smooth, line width=1.5pt] file {plotdata/stokes_9hullb.dat};
% 	\addplot[smooth, line width=1.5pt] file {plotdata/stokes_9hullc.dat};
% 	\addplot[smooth, line width=1.5pt] file {plotdata/stokes_9hulld.dat};
% 	\addplot[smooth, line width=1.5pt] file {plotdata/stokes_9hulle.dat};
% 	\addplot[mark=*, only marks, %
% 		mark options={%
% 		scale=1.3, fill=gray!30, draw=black, line width=0.8pt}] 
% 		coordinates {%
% 		(-1,0)
% 		(-0.8875,0)
% 		(-0.775,0)
% 		(-0.6625,0)
% 		(-0.55,0)
% 		(-0.4375,0)
% 		(-0.325,0)
% 		(-0.2125,0)
% 		(-0.1,0)
% 		(0,0)};
% 	\node at (axis cs:0.90,0.12)
% 	{\includegraphics[width=3.4cm]{figpdf/mini_9hull.pdf}};	
%   \end{axis}
% \end{tikzpicture}
% \endpgfgraphicnamed
\caption{From top to bottom: Stokes lines for the $2$-hull, Farrow and
Tuck's \citeyearpar{farrow_1995} $3$-hull, the $9$-hull, and the $6$-hull shown
before in Figure \ref{fig:nhull_Nhull}. For the $2$-hull and $6$-hull, the corner angles diverge at $\pm \pi/4$; for the remaining hulls, the corner angles are all rectangular. \label{fig:nhull_stokes}}
\end{figure}

As an example, consider Figure \ref{fig:nhull_stokes}, which illustrates the Stokes lines for various $N$-hulls, including a simple $2$-hull, the $3$-hull of \cite{farrow_1995}, a bulbous $6$-hull, and a step-like $9$-hull. With the exception of a single configuration, the condition that a Stokes line emerges into the upper-half plane is enough to guarantee that it intersects the free-surface. The exception is with the $3$-hull, for which the second singularity has a Stokes line emerging at an angle of $\nu_2 = 3\pi/5$, but which does not later encounter the free-surface.

To derive the form of the exponentials that appear whenever a Stokes Line
intersects the free-surface, we optimally truncate the asymptotic
expansions (\ref{eq:nhull_asympt}), and examine the remainder as the Stokes line is crossed. We let
\begin{equation} \label{eq:nhull_opttrun}
 q = \sum_{n=0}^{\mathcal{N}-1} \epsilon^n q_n + S_\mathcal{N},
\end{equation}

\noindent with a similar expression for the series for $\theta$. When $\mathcal{N}$ is chosen to be the optimal truncation point, the remainder $S_\mathcal{N}$ is found to be exponentially small, and by re-scaling near the Stokes line, it can be shown that a wave of the following form switches on:
\begin{equation} \label{eq:nhull_finalpre}
\sim \frac{2\pi i}{\epsilon^{\gamma_k}} 
 Q_k \exp\left[{-\frac{\chi_k}{\epsilon}}\right].
\end{equation}

\noindent To (\ref{eq:nhull_finalpre}), we must also include the complex conjugate due to the contributions from analytic continuation of the free-surface into the
lower-half $\phi$-plane [see (\ref{eq:nhull_bdint2})]. The sum of the two contributions is then
\begin{equation} \label{eq:nhull_final}
 q_{\text{exp}, k} \sim \minus \frac{4\pi}{\epsilon^{\gamma_k}} \Im \biggl\{ 
 Q_k \exp\left[{-\frac{\chi_k}{\epsilon}}\right] \biggr\},
\end{equation}

\noindent with, of course, one such expression for every $k\in \mathcal{J}$.

Thus, for any given arbitrary $N$-hull with a geometry such that $\mathcal{J}$
is nonempty, the appearance of exponentially small waves is a \emph{necessary
consequence} of the divergent low-Froude problem; in order to check that
such a ship can never be waveless, we need only verify that the sum of
all the contributions incurred can never be zero, so that there is a
non-zero wave amplitude far downstream. 

\subsection{Wave formulae for $N$-hulls} \label{sec:nhull_formula}

The constants $\gamma_k$ and $\Lambda_k$, which appear in the final form of the waves (\ref{eq:nhull_final}) [the latter as a prefactor in $Q_k$ in (\ref{eq:nhull_Q})], can be determined by re-scaling $w$ and $q$ near each of the singularities, and then matching the leading-order nonlinear (inner) solutions with the late-order (outer) terms of (\ref{eq:nhull_ansatz}). It can be shown (see (6.8) in \citealt{trinh_1hull}) that
\begin{equation} \label{eq:nhull_gammak}
\gamma_k = \frac{6\sigma_k}{1+3\sigma_k}
\end{equation}

\noindent and 
\begin{equation} \label{eq:nhull_lambdak}
\Lambda_k = \frac{ c_k^{6-3\gamma_k} e^{i\pi
\gamma_k/2}} {2C_k \left(1+3\sigma_k\right)^{\gamma_k}} 
\left[ \lim_{n\to\infty} \ \frac{\phi_{n, k}}{\Gamma(n+\gamma_k)} \right],
\end{equation}

\noindent where $C_k$ is given by
\begin{equation} \label{eq:nhull_C}
C_k = q_0^3(w^\bigstar) \exp\left(3i \int_{w^\bigstar}^{-a_k} \frac{\mathscr{H}\theta_1(\varphi)}{q_0^3(\varphi)}
 \ d\varphi \right).
\end{equation}

The terms $\phi_{n, k}$ are given by the recurrence relation, 
\begin{eqnarray}
 \phi_{0, k} &=& 1, \\
 \phi_{n, k} &=& \sum_{m=0}^{n-1} \left( m + \frac{2\sigma_k}{1 +
 3\sigma_k}\right) \phi_m \phi_{n-m-1} \text{\quad for $n \geq 1$.}
\end{eqnarray}

\noindent We will often make reference to the limiting ratio in (\ref{eq:nhull_lambdak}), so we define the function:
\begin{equation} \label{eq:nhull_omega_k}
\Omega(\sigma_k) \equiv \lim_{n\to\infty} \ \frac{\phi_{n,
k}}{\Gamma(n+\gamma_k)}.
\end{equation}

\noindent The value of $\Omega(\sigma_k)$ only depends on the local
divergence of the $k^\text{th}$ corner, and its values are given in
\cite{trinh_1hull}. Since $\Omega \neq 0$ for all choices of the local
angle $\sigma_k$, $\Lambda_k$ is also non-zero and this verifies that
each of the $|\mathcal{J}|$ corners of an $N$-hull must necessarily
generate a non-zero wave on the free surface.

With $Q_k$ given by (\ref{eq:nhull_Q}), $\Lambda_k$ given by (\ref{eq:nhull_lambdak}) and (\ref{eq:nhull_omega_k}), and $\text{arg}(c_k)$ from (\ref{eq:argck}), we have from (\ref{eq:nhull_final}) the result that
\begin{equation}
\begin{split}
q_{\text{exp}, k} \sim \minus \frac{4 \pi}{\epsilon^{\gamma_k}}
\frac{|c_k|^{6-3 \gamma_k} }{2 (1+3
  \sigma_k)^{\gamma_k}}  \frac{\Omega(\sigma_k)}{q_0^5} \exp\left[
-\Im\left( 3
 \int_{-a_k}^w 
  \frac{\mathscr{H}\theta_1}{q_0^3}\, {\rm d}{\phi}\right)\right]
\exp\left[-\frac{\Re(\chi_k)}{\epsilon}\right] \times \\ 
\cos\left[-\frac{\Im(\chi_k)}{\epsilon}
+\frac{\pi}{2}+ \frac{\pi \gamma_k}{2} + (6-3 \gamma_k)\theta_k + \Re\left( 3
   \int_{-a_k}^w 
  \frac{\mathscr{H}\theta_1}{q_0^3}\, {\rm
  d}{\phi}\right)\right].\label{eq:nhull_qexpk}
\end{split}
\end{equation}

%\begin{eqnarray}
% q_{\mbox{\scriptsize exp},k} &\sim& \frac{4 \pi}{\epsilon^{\gamma_k}}
%\frac{|c_k|^{6-3 \gamma_k} }{2 (1+3
%  \sigma_k)^{\gamma_k}}  \frac{\Omega(\sigma_k)}{q_0^5} \exp\left[
%-\Im\left( 3
% \int_{-a_k}^w 
%  \frac{\mathscr{H}[\theta_1]}{q_0^3}\, {\rm d}{\phi}\right)\right]
%\exp\left[-\frac{\Re(\chi_k)}{\epsilon}\right] \times  \nonumber
%\\ && \mbox{ } \cos\left[-\frac{\Im(\chi_k)}{\epsilon}
%+\frac{\pi}{2}+ \frac{\pi \gamma_k}{2} + (6-3 \gamma_k)\theta_k + \Re\left( 3
%   \int_{-a_k}^w 
%  \frac{\mathscr{H}[\theta_1]}{q_0^3}\, {\rm
%  d}{\phi}\right)\right].\label{eq:nhull_qexpk}
%\label{3.23}
%\end{eqnarray}

\noindent Then, for each $k \in {\cal J}$, we add the waves together, so that
the total wave contribution after all the Stokes lines have been
crossed is
\begin{equation}
q_{\mbox{\scriptsize exp}} \sim \sum_{k \in {\cal J}}
q_{\mbox{\scriptsize exp},k}. \label{eq:nhull_qexpsum}
\end{equation}

\section{The non-existence of waveless ships} \label{sec:nhull_nonexistence} 

\noindent Let us see what would be needed to produce a waveless ship. 

The wave contributions (\ref{eq:nhull_qexpk}) are written in terms of different denominators $\chi_k$ (also referred to as the `\emph{singulants}', \emph{c.f.} \citealt[p.148]{dingle_book}). To make it easier to sum them we rewrite them in terms of the single singulant, $\chi_1$. Note that 
\[ 
\chi_1(w) = {\rm i} \, \int_{-a_1}^{-a_k} \frac{{\rm d}{\phi}}{q_0^3} +
\chi_k(w),
\]

\noindent where in order for the integral to exist, we may have to avoid the intermediate corners by deforming the contour into the upper half plane. Consider now $q_{\text{exp}, k}$ in (\ref{eq:nhull_qexpk}) when $w$ is evaluated on the free-surface, that is, for $w \in \mathbb{R}^+$. From the appendix of \cite{trinh_1hull}, it was shown that 
\[ 
\exp\left[
-\Im\left( 3
 \int_{-a_1}^w 
  \frac{\mathscr{H}\theta_1}{q_0^3}\, {\rm d}{\phi}\right)\right] = q_0^3(w) \, {\rm e}.
\]

\noindent Moreover, the real part of $\chi_1(w)$ comes from the pole at infinity, giving
\begin{equation} \label{chi1res}
\Re(\chi_1) = 3 \pi \sum_{i=1}^N a_i \sigma_i.
\end{equation}

\noindent Putting these together in (\ref{eq:nhull_qexpk}) 
we find, for $w \in \mathbb{R}^+$, 
\begin{eqnarray*}
%\lefteqn{ 
q_{\text{exp}, k} \sim   \frac{\Lambda_k}{q_0^2(w)} 
\exp\biggl[-\frac{3 \pi}{\epsilon} \sum_{i=1}^N a_i \sigma_i \biggr]
% \times}&& \\ && \mbox{ } 
\cos\left[-\frac{\Im(\chi_1(w))}{\epsilon}+
 \Re\left( 3 \int_{-a_1}^w 
  \frac{\mathscr{H}\theta_1}{q_0^3}\, {\rm d}{\phi}\right) +\Psi_k
 \right],
\end{eqnarray*}
where the dependence on $k$ arises only through the constants
\begin{eqnarray*}
 \Lambda_k &=& \minus \frac{4 \pi{\rm e}}{\epsilon^{\gamma_k}}
\frac{|c_k|^{6-3 \gamma_k} }{2 (1+3
  \sigma_k)^{\gamma_k}}\Omega(\sigma_k)\exp\left[
\Im\left( 3
 \int_{-a_1}^{-a_k} 
  \frac{\mathscr{H}\theta_1}{q_0^3}\, {\rm d}{\phi}\right)\right]
\exp\left[-\frac{1}{\epsilon}\,\Im\left(\int_{-a_1}^{-a_k}
    \frac{{\rm d}{\phi}}{q_0^3}\right) \right],\\
\Psi_k & = &\frac{1}{\epsilon}\,\Re\left( \int_{-a_1}^{-a_k}\frac{{\rm d}{\phi}}{q_0^3}
\right)-
 \Re\left( 3 \int_{-a_1}^{-a_k} 
  \frac{\mathscr{H}\theta_1}{q_0^3}\, {\rm d}{\phi}\right)
+\frac{\pi}{2}+ \frac{\pi \gamma_k}{2} + (6-3 \gamma_k)\theta_k.
\end{eqnarray*}

\subsection{On the two-cornered ship ($2$-hull)}

We have already shown in \cite{trinh_1hull} that a single-cornered hull must produce waves.  Therefore let us consider first the next simplest case of a 2-hull. For such a ship to be waveless, both corners must generate Stokes lines which intersect the free surface, and the waves generated by each must exactly cancel; then, there will be a finite section of free surface containing waves, but no wavetrain at infinity (as happens in the case of capillary waves in \citealt{chapman_2002}).

In order for the waves from the two corners to cancel, we need $\Lambda_{1} = \Lambda_{2}$. Now for a fixed value of $\epsilon$ this may indeed be possible (and we give such an example in \S\ref{sec:nhull_numerical}), but what if we want the waves to vanish for {\em all} (small) values of $\epsilon$? Then, since each $\Lambda_k$ is of the form 
\[ 
\lambda_1 \, \epsilon^{\lambda_2} \,{\rm e}^{- \lambda_3/\epsilon},
\]

\noindent as $\epsilon \rightarrow 0$, we need the exponentials to be equal, the powers of $\epsilon$ to be equal, and the prefactors to be equal. In order for the exponentials to be equal, we require
\begin{equation} \label{1}
\Im\left(\int_{-a_{1}}^{-a_{2}}
    \frac{{\rm d}{\phi}}{q_0^3}\right)=0,
\end{equation}

\noindent and since
\[ 
\arg \left(\frac{1}{q_0^3}\right) = - 3  \theta_k\quad
\mbox{ for }a_k<w<a_{k+1}
\]

\noindent the only way (\ref{1}) can hold is if $\theta_1 = \pi/3$, so that $q_0^3$ is real for $-a_{1}<w<-a_{2}$. Now, for the algebraic factors of $\epsilon$ to be equal, we require $\gamma_1 = \gamma_2$, which implies $\sigma_1=\sigma_2$. Thus the only possibility for a waveless 2-hull is for a ship with $\sigma_1 = \sigma_2 = 1/3$. To eliminate this final possibility we need to consider the prefactors. Since
\[ 
\Im\left(\mathscr{H}\theta_1\right) = \left\{ \begin{array}{ll}
0 & w<0,\\
\theta_1(w) & w>0,
\end{array}
\right.
\]

\noindent and $q_0^3$ is real for $-a_{1}<w<-a_{2}$, then
\[
\Im\left( 3
 \int_{-a_1}^{-a_2} 
  \frac{\mathscr{H}\theta_1}{q_0^3}\, {\rm d}{\phi}\right) = 0.
\]

\noindent The only remaining difference between the two prefactors is in
$c_k$. However, since
\[ c_1^3 = \frac{a_1^{2}}{(a_1-a_2)},
\qquad
c_2^3 = \frac{a_2^{2}}{(a_1-a_2)},\]
the only way that we can have $|c_1| = |c_2|$ is if $a_1 = a_2$.
Thus the two prefactors must be different, one corner always
dominates the other one, and the waves can 
never cancel. Waveless ships with two corners are not possible.

\subsection{On general $N$-cornered ships ($N$-hulls)}

What can we say about more general ships? Let us take a general $N$-hull with the following assumptions: suppose that all the Stokes lines intersect the free surface, that $\sigma_k>0$ for each $k$, and that $\theta_N\leq 2 \pi/3$. In Figure 1, hulls (a) to (e) satisfy this requirement, whereas hulls (f) to (h) do not. Under these assumptions, $\theta_k$ is monotonically increasing with $k$, so that  $\arg(1/q_0^3)$ is monotonically decreasing. Thus the argument of the exponential 
\[ 
-\frac{1}{\epsilon}\,\Im\left(\int_{-a_1}^{-a_k}
    \frac{{\rm d}{\phi}}{q_0^3}\right) 
\]

\noindent is convex in $k$, increasing while $0<\theta_k<\pi/3$ and then decreasing  while $\pi/3<\theta_k<2\pi/3$. Thus if $\theta_j \not = \pi/3$ for all $j$ then we can see immediately that the waves generated at the corner $k$ such that $\theta_{k-1}<\pi/3<\theta_k$ exponentially dominate all the others. On the other hand, if $\theta_k  = \pi/3$ then  the waves from corners $k$ and $k+1$ have the same exponential factor (as in the 2-hull case). If we further impose $\sigma_k = \sigma_{k+1}$ then they have the same algebraic factor, and there is possibility of wave cancellation if the prefactors are equal.

Of course, even if we could get the prefactors to be equal, we still have to worry about the waves generated by all the other corners. In fact, even from these two corners there would be higher-order correction terms (both in the form $\epsilon {\rm e}^{-c/\epsilon}$, $\epsilon^2 {\rm e}^{-c/\epsilon}$, \emph{etc.} and also in the form of a trans-series ${\rm e}^{-c/\epsilon}$, ${\rm e}^{-2c/\epsilon}$ \emph{etc.}). Thus it does not seem to be worth pursuing the analysis further. However, even if we cannot get the waves to cancel exactly, we might expect a significant reduction in the amplitude of the waves in the case when leading-order cancellation occurs.

This brings us to the natural question of whether the analysis we have presented may be used to design a hull to minimise the wave drag. Before we address this question, let us first demonstrate that the hulls shown in Figure 1 (f) through (h) must generate waves on the free surface.

We consider them in reverse order. The 9-hull shown in Figure 1(h) has $|{\cal J}|=5$ (as shown in Figure 3), with $\arg(1/q_0^3)$ alternating between zero and $-3 \pi/2$; thus the argument above can be used to show that the
contribution from $a_1$ exponentially dominates all the others. The hull shown in Figure 1(g) has $|{\cal J}|=4$, with three positive angles $\sigma$ and one negative angle. Thus the algebraic factor in the contribution from $w=-a_5$ is different to the others, and those waves must always exist on the free surface.

Finally let us consider the 3-hull shown in Figure 1(f), which is
found in the work of \cite{farrow_1995}, and for which the addition
of a downward pointing bulb was shown to dramatically reduce the wave
resistance compared to a rectangular ship. In this case, $|{\cal J}|=1$,
and $w=-a_3$ is the only relevant corner, so there are always
waves on the free surface. The principal effect of the bulb is to
lower the usual singularity farther away from the free surface,
thereby decreasing the amplitude of the waves.

This last example not only highlights the difficulty in trying to minimise the wave drag, but also the advantage of our semi-analytic approach: we have gained considerable insight into the mechanism of wave production; from this, we can immediately see why Farrow \& Tuck obtained the results that they did.

In trying to design reduced-wave hulls, it is crucial to specify what exactly is the optimisation process. For example, if we simplify the hull of Farrow \& Tuck to a 2-hull by reducing the width of the downward pointing bulb to zero, then we have one parameter, $a_1$, over which we can optimise (since $a_1 + a_2=1$). We find the smallest waves occur for $a_1=0.5$, \emph{i.e.} when there is no bulb. However, in our current nondimensionalisation, as we vary $a_1$, both the depth of the hull and that of the bulb vary. If instead we fix the depth of the hull, and allow the depth of the bulb to vary, we find that the waves get smaller as the bulb gets deeper. On the other hand, if we fix the depth of the bulb, and allow the depth of the hull to vary, we again find that the smallest waves correspond to the hull depth being equal to the bulb depth, \emph{i.e.} to there being no bulb.

\section{Numerical and asymptotic results for two-cornered hulls} \label{sec:nhull_numerical}

%\noindent In Part 1, we presented two algorithms which can be used to perform the numerical computations of the nonlinear stern problem (\ref{eq:nhull_bdint2})--(\ref{eq:nhull_bern2}), particularly at small values of $\epsilon$. 

\noindent The numerical algorithms developed in \cite{trinh_1hull} can be used to verify the asymptotic predictions. Here, we focus on the particular case of a $2$-hull, which we refer to as a $[\sigma_1, \sigma_2]$-hull; this is a ship with divergent corner-angles $\sigma_1$ and $\sigma_2$, and with leading-order flow given by (\ref{eq:nhull_rigid}), or
\begin{equation} \label{eq:nhull_2hull}
q_s = \frac{w^{\sigma_1+\sigma_2}}{(w+a_1)^{\sigma_1} (w+a_2)^{\sigma_2}},
\end{equation}

\noindent with $a_1 + a_2 = 1$. 

\begin{figure}\centering
\includegraphics[width=1.0\linewidth]{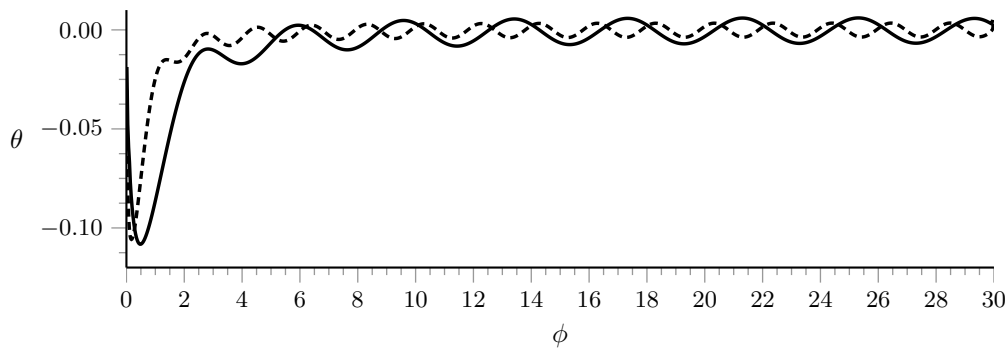}
% \begin{preview}
%  \begin{tikzpicture} 
%  \begin{axis}[
%  	xlabel=$\phi$, ylabel=$\theta$,
%  	xmin=0,xmax=30, ymin=-0.12, ymax=0.01,
%  	width=0.97\textwidth, height=5cm,
%  	yticklabel style= {/pgf/number format/fixed, 
%  			/pgf/number format/fixed zerofill, 
%  			/pgf/number format/precision=2}, 
%  	scaled ticks=false] 
%  	\addplot[densely dashed, line width=1.3pt] file {plotdata/solo2525.dat};
%    	\addplot[line width=1.3pt] file {plotdata/solo50125.dat};
%  \end{axis}	
%  \end{tikzpicture}
% \end{preview}
\caption{Solutions for the $[0.5, 0.125]$-hull (dashed line) and $[0.25,
0.25]$-hull (solid line) at $\epsilon = 2/3$ and $\epsilon = 1/3$, respectively. Both
ships have corners set at $a_1 = 0.8$ and $a_2 = 0.2$. The solutions were computed using \textsc{algorithm a} of \cite{trinh_1hull} with $n = 1000$ and $\Delta\phi = 0.015$ for the former ship and $n = 2000$ and $\Delta\phi = 0.015$ for the latter. \label{fig:nhull_specsol}} 
\end{figure}

As we discussed in \cite{trinh_1hull}, the numerical computation of the stern problem at small values of $\epsilon$ can be particularly difficult, and the culprit is the presence of the attachment singularity at $w = 0$, associated with a small boundary layer; this singularity is responsible for most of the numerical error. For hulls where the in-fluid attachment angle between the free-surface and body is less than $\pi/3$, a simple finite difference scheme based on the methods outlined in \cite{vb_1977} can be used, provided that we limit our search to waves larger than $\approx 10^{-4}$. Figure \ref{fig:nhull_specsol} provides an example of solutions found using this method. 

\begin{figure}
\includegraphics[width=1.0\linewidth]{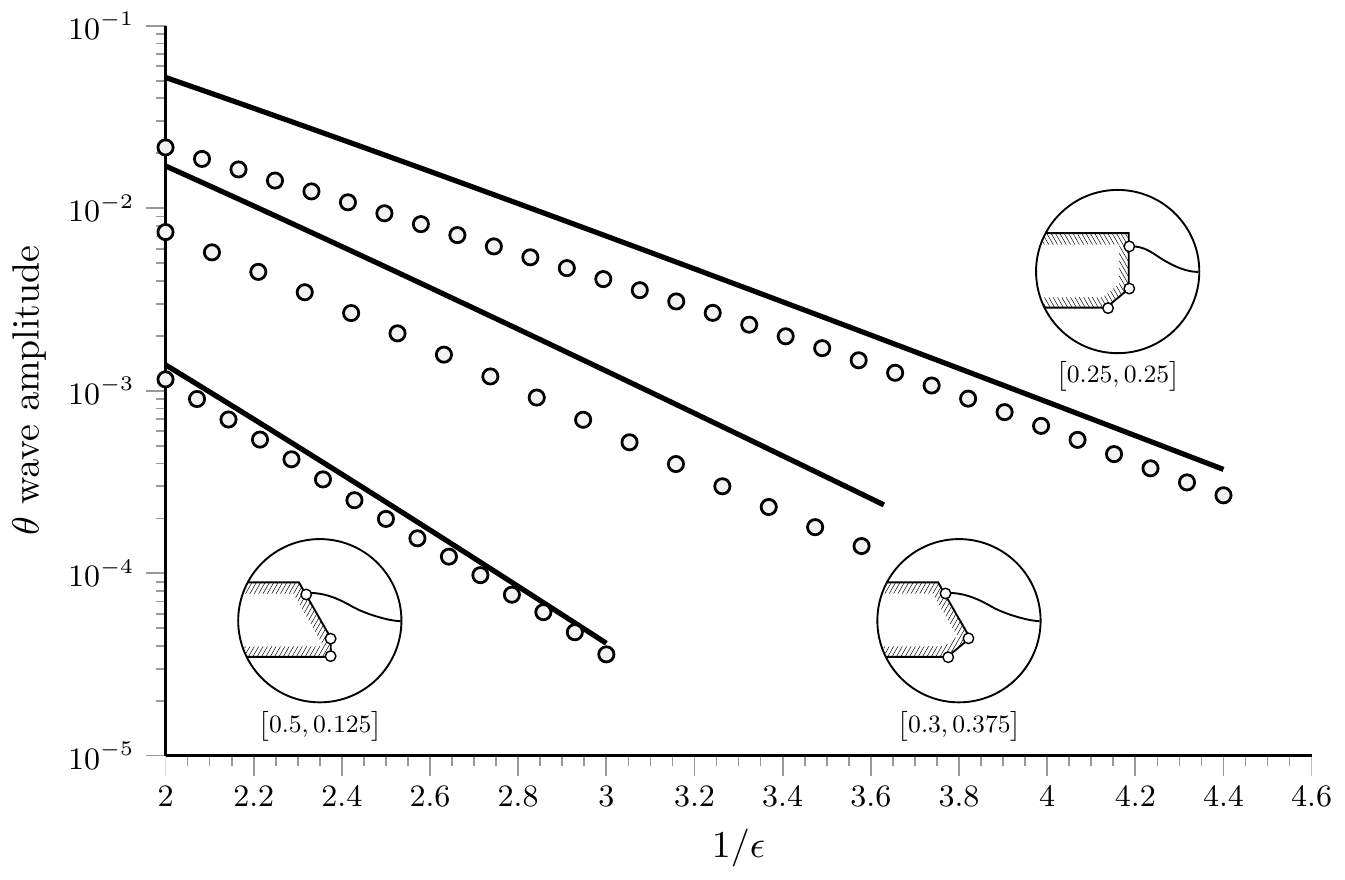}
  \caption{Numerical (dots) and asymptotic (solid) amplitudes of the
    downstream waves  for a range of hull inclinations. In all cases,
    the corner points are fixed with $a_1 = 0.8$ and $a_2 = 0.2$. The solutions were computed using \textsc{algorithm a} of \cite{trinh_1hull}. The parameters used were the following: \ding{172} $n = 1000$, $\Delta\phi = 0.04$; \ding{173} $n = 1500$, $\Delta\phi = 0.03$; and \ding{174} $n = 2000$, $\Delta\phi = 0.025$. \label{fig:nhull_num_final}}
\end{figure}

The theory of \S\ref{sec:nhull_asymptotic} and \S\ref{sec:nhull_nonexistence} can be verified by comparing the analytical predictions with numerically computed wave amplitudes far from the ship. First, consider the effect of varying the Froude number on ships of a fixed geometry. This is shown in Figure \ref{fig:nhull_num_final} for three $2$-hulls with their corners fixed with $a_1 = 0.8$ and $a_2 = 0.2$. The individual cosine waves are calculated from (\ref{eq:nhull_qexpk}) with $q_0 \to 1$ downstream, and then the final amplitude computed using the sum (\ref{eq:nhull_qexpsum}) (see \S4.5 in \citealt{trinh_thesis} for additional details). The match between numerical and asymptotic solutions is quite good, and like the previous work, we remark that the results are applicable over a wide range of Froude numbers.  

\begin{figure} \centering
\includegraphics[width=1.0\linewidth]{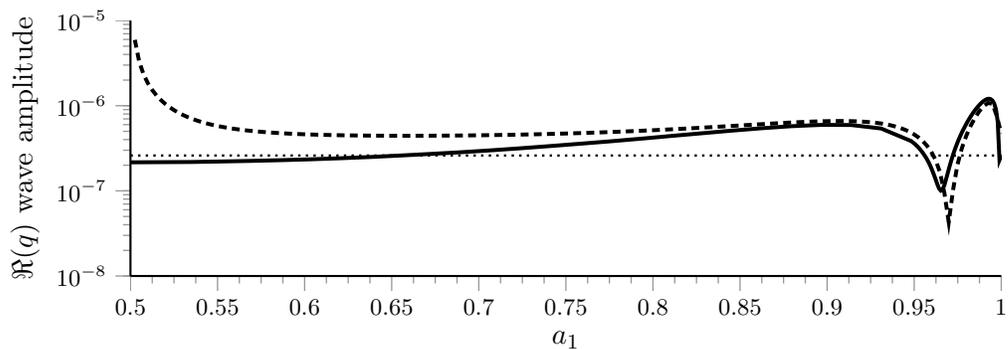}
% \begin{preview}
%  \begin{tikzpicture}  
%    \begin{semilogyaxis}[xlabel=$a_1$, 
%  		ylabel={$\Re(q)$ wave amplitude}, 
%  		ylabel style={rotate=90, font=\large},
%  		xlabel style={font=\large},
%  		ymin=0.00000001, ymax=0.00001,
%  		width=0.98\textwidth, height=5cm]
 		
%  	\addplot[line width=1.5pt] 
%  		file {plotdata/cc2525_ep015_amp.dat};
%  	\addplot[densely dashed, line width=1.5pt] file
%  			{plotdata/cc2525_ep015_2corn.dat};
%  	\draw[dotted, line width=0.9pt] 
%  		(axis cs:0.5, 2.6E-7) -- (axis cs:1.0,2.6E-7);
%    \end{semilogyaxis}
%    \end{tikzpicture}
% \end{preview}
\caption{The numerical solution (solid) is plotted against the asymptotic approximation (dashed) for the simplified nonlinear problem of \S\ref{sec:cc_simpnon}. The ship is a $[\tfrac{1}{4}, \tfrac{1}{4}]$-hull with $a_1 + a_2 = 1$ and $\epsilon = 0.15$. The dotted line indicates the one-cornered approximation for a rectangular hull. \label{fig:nhull_cc2525.1}}
\end{figure}

%The two-cornered approximation very accurately predicts the solution when the two corners are well separated, but is singular when $a_1, a_2 \approx 0.5$ near the left.  

%In Figure \ref{fig:nhull_cc2525.1} we show the effect of varying the
%position of the corners by showing the wave amplitude for the 
% $[\tfrac{1}{4}, \tfrac{1}{4}]$-hull with $\epsilon=0.15$ as a function of $a_1$.
%This figure shows a number of interesting effects. The first is that 
%the numerically computed wave amplitude shows a significant dip (an
%order of magnitude) near
%$a_1 = 0.96$, and that this is captured fairly accurately in the asymptotic
%solution.

Next, we would like to consider the effect of fixing the Froude number, but varying the positions of the corners. The problem, however, is that the interesting effects of this procedure are only seen at values of $\epsilon$ much smaller than what we can achieve using the above numerical methods. In Appendix \ref{sec:cc_simpnon}, we present a slightly simplified version of the ship-wave problem (\ref{eq:nhull_bdint2})--(\ref{eq:nhull_bern2}) that preserves the asymptotic structure of the waves, but also enables us to compute numerical solutions to much higher accuracy. 

This simplified problem was used to create Figure \ref{fig:nhull_cc2525.1}, which shows the effect of varying the positions of the corners on the wave amplitude for a $[\tfrac{1}{4}, \tfrac{1}{4}]$-hull with $\epsilon = 0.15$, $a_1 + a_2 = 1$, and values of $0.5 \leq a_1 \leq 1$. The figure contains a number of interesting effects: the first is that the numerically computed wave amplitude shows a significant dip (an order of magnitude) near $a_1 = 0.96$, and that this effect is also captured fairly accurately in the asymptotic solution.

The reason for the dip is that at this value of $a_1$, the waves from the two corners exhibit partial destructive interference. However, from the of the previous section, we know that the waves generated by the corner at $-a_2$ should exponentially dominate those from $-a_1$. How then are they cancelling each other? The answer is that the prefactor $|c_1|^{6 - 3 \gamma_1}$ is over ten times larger than $|c_2|^{6  - 3 \gamma_1}$; at this particular value of $\epsilon$, the difference in prefactors is enough to compensate for the difference in exponentials, since 
\[ 
\Im \left(\int_{-a_1}^{-a_2} \frac{{\rm d} \phi}{q_0^3} \right)
\]

\noindent is not very large. Thus, the value of $a_1$ at which cancellation occurs depends on $\epsilon$; for somewhat smaller of $\epsilon$ the Stokes line from $w = -a_2$ does indeed dominate and no cancellation is possible. This indicates that it should be possible to design hulls with reduced wave drag at a particular speed (Froude number). It is also reassuring to note that including the leading-order term from each of our exponentially-small waves captures the behaviour of the solution very well, even though formally, one of the terms is exponentially subdominant.

The second effect illustrated by Figure \ref{fig:nhull_cc2525.1} is the divergence between our asymptotic expansion and the numerical solution for values of $a_1$ close to $0.5$. The reason for the divergence is that the prefactors $c_1$ and $c_2$ are singular as the corners approach each other; our analysis in \S\ref{sec:nhull_asymptotic}, in fact, implicitly assumes that the corners of the ship are spaced sufficiently far from one another.  

To be more specific, in order to determine the constants $\gamma_k$ and $\Lambda_k$ in (\ref{eq:nhull_gammak}) and (\ref{eq:nhull_lambdak}) in the previous analysis, the outer solution was required to match a nonlinear inner solution. The size of this inner region can be derived by observing where the breakdown in the outer expansion (\ref{eq:nhull_asympt}) first occurs, \emph{i.e.} where $\epsilon q_1 \sim q_0$. From (\ref{eq:nhull_rigid2}) and (\ref{eq:nhull_q1}), the required re-scaling of $w$ near a singularity at $-a_k$ can be seen to be
\begin{equation} \label{eq:nhull_innerscaling}
 w+a_k = \mathcal{O}\left(\epsilon^{\frac{1}{1+3\sigma}}\right).
\end{equation}

\noindent Thus, if two (or more) singularities are spaced within a distance of (\ref{eq:nhull_innerscaling}) apart, then the previous asymptotic methodology breaks down as $\epsilon \to 0$. If we combine the two corners into a single corner of angle $\pi/2$, we find the wave amplitude is given by the dotted line in Figure \ref{fig:nhull_cc2525.1}. This approximation clearly works well for $a_1$ close to 0.5. 

A uniform approximation would need to smoothly match with the one-cornered approximation at one end, and the (separated) two-cornered analysis at the other, and thus bridges the dashed and dotted lines in Figure \ref{fig:nhull_cc2525.1}. Such an approximation requires us to consider the distinguished limit in which the corners of the ship are allowed to approach one another as $\epsilon \to 0$. This is similar to the situation in \cite[Appendix B]{chapman_2006} where the asymptotic solutions for flow over a rectangular step in a channel was analysed in the case where both corners of the step lie in the same `inner region'. For the case of the multi-cornered ship, the details of this analysis are very technical, and will be published in a future paper. 

\section{Discussion} \label{sec:discuss}

\noindent \emph{In the end, what is the definitive answer to the conundrum of existence and non-existence of waveless ships?} Unlike our results for the single-cornered ships of \cite{trinh_1hull}, there does not seem to be a simple answer to this question, applicable to the most general piecewise-linear hulls. Despite this, however, we have offered several new insights into the study of ship-wave resistance: we have offered explicit formulae for the computation of waves given the shape of the ship's hull; we have offered simple interpretations of the production of such waves in terms of Stokes line crossings and the Stokes Phenomenon; and perhaps most importantly, we offered a methodology which, given specific ships, provides an immediate and intuitive understanding of the effect of the body on the free-surface.

In the previous work, we highlighted the importance of distinguishing between local and global properties of the analysis. Consider the factorial over power divergence of the asymptotic series in (\ref{eq:nhull_ansatz}), or the emergence-conditions of Stokes lines in (\ref{eq:nhull_stokescond}), or the numerically-determined pre-factor, $\Omega_k$, in (\ref{eq:nhull_omega_k})---these are all \emph{local} properties of the problem; indeed their derivation only depends on the behaviour of the asymptotic solution near the relevant singularities. These local properties, we understand well. 

In contrast, many \emph{global} questions remain unanswered. For example, given a ship, represented by $q_0$, what are the necessary and sufficient conditions for Stokes lines to intersect the free surface? Or perhaps more difficult: what are the necessary restrictions on $q_0$ so that total phase cancellation occurs? We have provided a few preliminary results on this global problem, but a more exhaustive analysis of these issues remains an open problem. 

Naturally, our study of ship waves would be incomplete without a theory applicable for smooth hulls, with the eventual goal of addressing the well-known technique of using a bulb to reduce the wave resistance of a ship \citep{baba_1976}. However, the difficulty here is that analytic continuation is an ill-posed process and small perturbations in the shape of a hull can have large effects on the associated singularities---a unified theory for arbitrary ship geometries will likely prove difficult, if not downright impossible.

\begin{figure}
\includegraphics[width=1.0\linewidth]{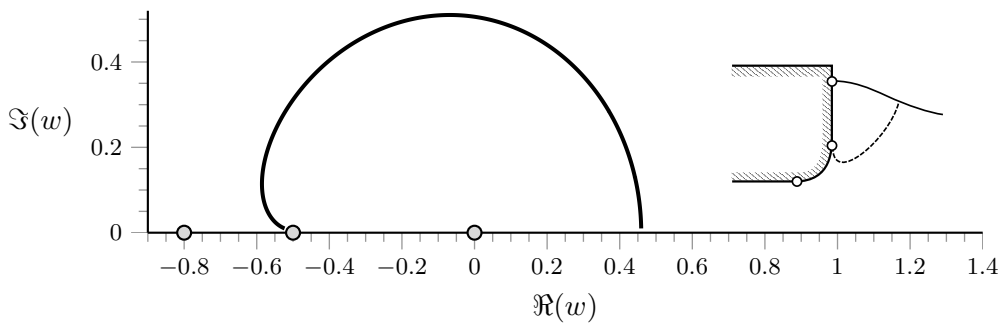}
% \beginpgfgraphicnamed{nhull_stokessm}
% \begin{tikzpicture}  
%   \begin{axis}[ xlabel=$\Re(w)$, 
% 		ylabel=$\Im(w)$, 
% 		ylabel style={font=\large},
% 		xlabel style={font=\large},
% 		ymin=0, ymax=0.52,
% 		xmin=-0.9, xmax=1.4,
% 		%xtick={0.0}, ytick=\empty,
% 		%major tick length=0.22cm,
% 		%minor tick length=0.0cm,
% 		width=0.93\textwidth, height=4.5cm]
% 		
% 	\addplot[smooth, line width=1.5pt] file {plotdata/stokes_smhull.dat};
% 	\addplot[mark=*, only marks, %
% 		mark options={%
% 		scale=1.3, fill=gray!30, draw=black, line width=0.8pt}] 
% 		coordinates {(-0.8,0) (-0.5,0) (0,0)};
% 	\node at (axis cs:1,0.25)
% 	{\includegraphics[width=2.8cm]{figpdf/mini_sm.pdf}};	
%   \end{axis}
% \end{tikzpicture}
% \endpgfgraphicnamed
\caption{%
Stokes lines for the smoothed hull in (\ref{eq:remarks_thetaship}) with $a_1 = 0.8$ and $a_2 = 0.5$. The Stokes line leaves tangentially along the boundary, but later intersects the free-surface. \label{fig:remarks_smhull}}
\end{figure}

Perhaps, then, we should only consider specific classes of smooth ships. Ship waves associated with continuous geometries have been considered in the numerical work of \cite{tuck_1984}, \cite{madurasinghe_1988}, and \cite{farrow_1995}, where there, the hulls are specified by piecewise-entire functions. For example, \cite{farrow_1995} consider the family of hulls given by
\[
\theta = \begin{cases}
0 & \text{for $w\in(-\infty,-1)$} \\
A(w+1)(w+b) + \frac{\pi}{2}\frac{(w+1)}{(1-b)} & \text{for $w\in(-1,-b)$} \\
\frac{\pi}{2} & \text{for $w\in(-b,0)$}
\end{cases}
\]

\noindent which, given parameters $A$ and $b$, provides a ship consisting of a horizontal bottom and a vertical line, joined by a rounded section; $A > 0$ yields rounded corners and $A < 0$ yields bulbous sterns. The key is that if we restrict ourselves to classes of ships given by piecewise entire functions, then the complex singularities must be located at the points joining each piece. As a simpler example, we may consider the ship with
\begin{equation} \label{eq:remarks_thetaship}
\theta = \begin{cases}
0 & \text{for $w\in(-\infty,-a_1)$} \\
\frac{\pi}{2}\left[ 1 + \frac{(w +a_2)}{(a_1-a_2)}\right] & \text{for $w\in(-a_1,-a_2)$} \\
\frac{\pi}{2} & \text{for $w\in(-a_2,0)$},
\end{cases}
\end{equation}

\noindent which is similar to the vertically-faced one-cornered ships studied previously, but with a rounded edge. Analysis of the Stokes lines shows that the relevant line emerges from $w  = -a_2$; this is shown in Figure \ref{fig:remarks_smhull}. The study of these piecewise-entire ships is the subject of ongoing investigation. 

A similar direction for research is towards the development of a low-Froude asymptotic theory for flows past three-dimensional, full-bodied ships. This builds upon the works of, for example, \cite{keller_1979} and \cite{brandsma_1985}, who apply geometrical ray theory to the case of streamline (thin) ships. In theory, the interpretation we have presented in this paper of free-surface waves arising due to Stokes line crossings is still valid in three dimensions, except now, singularities are associated with Stokes \emph{surfaces} rather than \emph{lines}. In practice, however, the analysis is complicated due to the loss of complex variable techniques. We refer the reader to the work of \cite{chapman_2005}, which provides a first step towards extensions of exponential asymptotics to partial differential equations.

In addition to our study, which solely focuses on the low-Froude model of \cite{dagan_1972}, it is important for us to question the place of these simplified mathematical models in terms of the bigger picture: that which includes the effects of vorticity, viscosity, and time-dependence in ship-wave interactions. As we elucidated in the introduction, numerical work (as particular examples, see \citealt{grosenbaugh_1989} and \citealt{yeung_1997}) show that in practice, these neglected effects can have significant roles in the production of waves. Extended discussions of the role of low-Froude theories appear in \citeauthor{tuck_1984} (\citeyear[p. 301]{tuck_1984}), \cite{tuck_1991}, and \cite{tulin_2005}. Thus, while it is certainly true that in order to obtain \emph{analytical} approximations directly relating ship geometries to free-surface waves, the low-Froude approximation provides enormous simplification, we hope that similar analytical theories can be developed which include a more complete host of effects.

\bibliographystyle{jfm}
%\bibliography{/Users/trinh/work/documents/bib/philmaster}
\providecommand{\noopsort}[1]{}

\appendix

\section{The simplified non-linear problem} \label{sec:cc_simpnon}

\noindent The full problem in (\ref{eq:nhull_bdint2})--(\ref{eq:nhull_bern2}) can be studied using the methods we develop here, but can also work with a simpler problem that nevertheless contains all of the the key components. The reason for this simplification is that, in order to verify the asymptotic analysis in the regime where the ship's corners are closely spaced, wave amplitudes must be computed to five or six digits of precision---otherwise, the fine effects of adjusting the ship's geometry are easily missed; this precision can only be easily achieved for the simpler problem, which we now derive.

As we know, when exponentially small terms are sought from (\ref{eq:nhull_bdint2}), the integral term, $\mathscr{H}\theta$, only serves a minor role throughout the analysis. If we return to the derivation of the late-orders
ansatz (\ref{eq:nhull_ansatz}), we recall that the subdominance of $\mathscr{H}\theta$ as $n \to \infty$ ensures that it plays no part in the derivation of $\chi_k$. In fact, the only role of the Hilbert Transform is to change the expression for $q_1$ in (\ref{eq:nhull_q1}). Consequently, in the final form of the waves $q_{\text{exp}, k}$ in (\ref{eq:nhull_qexpk}), the presence of $\mathscr{H}\theta_1$ only serves to change the amplitude coefficient and the phase shift by an $\mathcal{O}(1)$ amount. 

Therefore, the salient features of the problem can still be retained if we use $\log q \pm i\theta = \log q_s$ instead of (\ref{eq:nhull_bdint2}); this way, we simplify the full problem in (\ref{eq:nhull_bdint2}) to (\ref{eq:nhull_bern2}) to a single nonlinear differential equation in $q$. Analytic continuation into the upper-half plane, and substituting $i\theta = \log(q_s/q)$ into (\ref{eq:nhull_bern2}) gives
\begin{equation} \label{eq:nhull_simpbern}
\epsilon q_s q^3 \frac{dq}{dw} + \frac{i}{2} \Bigl[q^2 - q_s^2\Bigr] = 0,
\end{equation}

\noindent which can be solved subject to the single boundary condition $q(0) = 0$. It is more convenient to work under the substitution $\eu(w) = q^2(w)$, where we have
\begin{equation} \label{eq:nhull_simpbernphi}
\epsilon q_s \eu \frac{d\eu}{dw} + i \Bigl[ \eu - q_s^2 \Bigr] = 0,
\end{equation}

\noindent as a simplified nonlinear model of ship waves. Simplifications of the boundary integral problem (\ref{eq:nhull_bdint2}) and (\ref{eq:nhull_bern2}) were also proposed in \cite{tuck_1991, tuck_1991b}, but there, the simplifications were argued based on behavioural requirements. Here, (\ref{eq:nhull_simpbern}) is a justified reduction based on the $\epsilon \to 0$ limit.

Notice that in this new problem, we chose to analytically continue into the upper half-$w$-plane, and thus the exponentially small waves of (\ref{eq:nhull_simpbern}) will possess both a real and imaginary part. If we wish, we can mirror the analysis for the lower half-$w$-plane and add the complex conjugate as we did before for (\ref{eq:nhull_final}). 

However, it is somewhat simpler to examine (\ref{eq:nhull_simpbern}) as a problem on its own; thus we shall only concern ourselves with studying the real component of the solution to (\ref{eq:nhull_simpbern}), which we write $\ol{q}_\text{exp} = \Re(q_\text{exp})$. Now instead of (\ref{eq:nhull_qexpk}), the form of the waves for the
simplified problem (with well-separated corners) is given by 
\begin{multline}\label{eq:nhull_cc_qexpk}
\ol{q}_{\text{exp}, k} \sim \minus \frac{2 \pi}{\epsilon^{\gamma_k}}
\frac{|c_k|^{6-3 \gamma_k} }{2 (1+3
  \sigma_k)^{\gamma_k}}  \frac{\Omega(\sigma_k)}{q_0^5} 
\exp\left[-\frac{\Re(\chi_k)}{\epsilon}\right] \times \\ 
\cos\left[-\frac{\Im(\chi_k)}{\epsilon} + \frac{\pi \gamma_k}{2} + (6-3 \gamma_k)\theta_k \right],
\end{multline}

\noindent which is effectively (\ref{eq:nhull_qexpk}) with $\mathscr{H} \equiv 0$ and without a phase shift of $\pi/2$. The reduction by a factor of $2$ in (\ref{eq:nhull_cc_qexpk}) compared to (\ref{eq:nhull_qexpk}) occurs because there is no need to add the complex conjugate wave contribution. Analytical and numerical results for the simplified nonlinear problem of (\ref{eq:nhull_simpbern}) in the context of a one-cornered ship can be found in \cite{trinh_1hull}, whereas we have already discussed the numerical solution of the $[\tfrac{1}{4}, \tfrac{1}{4}]$-hull for the simplified problem in \S\ref{sec:nhull_numerical} and Figure \ref{fig:nhull_cc2525.1}. 
\end{document}